\documentclass[prb,superscriptaddress,reprint,twocolumn,amssymb,aps,showkeys]{revtex4-1}
\usepackage{amsmath}%
\usepackage{amsfonts}%
\usepackage{amssymb}%
\usepackage{graphicx}
\usepackage{color}
\usepackage{tabularx}
\usepackage{braket}
\usepackage{mathrsfs}

\begin{document}
\title{Strong Coupling between Whispering Gallery Modes and Chromium Ions in Ruby}

\author{Warrick G. Farr}
\affiliation{ARC Centre of Excellence for Engineered Quantum Systems, University of Western Australia, 35 Stirling Highway, Crawley WA 6009, Australia}

\author{Maxim Goryachev}
\affiliation{ARC Centre of Excellence for Engineered Quantum Systems, University of Western Australia, 35 Stirling Highway, Crawley WA 6009, Australia}

\author{Daniel L. Creedon}
\affiliation{ARC Centre of Excellence for Engineered Quantum Systems, University of Western Australia, 35 Stirling Highway, Crawley WA 6009, Australia}

\author{Michael E. Tobar}
\affiliation{ARC Centre of Excellence for Engineered Quantum Systems, University of Western Australia, 35 Stirling Highway, Crawley WA 6009, Australia}


\begin{abstract}
We report the study of interactions between cavity photons and paramagnetic Cr$^{3+}$ spins in a ruby (Cr$^{3+}$:Al$_2$O$_3$) Whispering Gallery mode (WGM) resonator. Examining the system at microwave frequencies and millikelvin temperatures, spin-photon couplings up to 610 MHz or about $5$\% of photon energy are observed between the impurity spins and high quality factor ($Q > 10^5$) WGM. 
Large tunability and spin-spin interaction allows operation in the strong coupling regime. The system exhibits behaviour not predicted by the usual Tavis-Cummings model because of interactions within the two-level spin bath, and the existence of numerous photonic modes. 

\end{abstract}
\date{\today}
\maketitle

\section*{Introduction}

Quantum strong coupling regimes have attracted interest as entangled states necessary for quantum computing schemes. The Tavis-Cummings model (TCM)\cite{PhysRev.170.379,PhysRev.188.692,PhysRevLett.103.043603} describes the coupling between \color{black} an ensemble of \color{black} two level systems (TLS) and a quantized mode of an electromagnetic field, and forms the basis of the large areas of research known as Cavity and Circuit Quantum Electodynamics (QED). The strong coupling regime at microwave frequencies has been demonstrated in a number of experiments (nitrogen-vacancy (NV) centres in diamonds\cite{ROOMNV06}, waveguides\cite{Teufel2011}, superconducting qubits\cite{Wallraff2004}, rare earth ions\cite{PhysRevLett.110.157001}, and ultra-cold atoms\cite{PhysRevA.82.033810}). However, the model does not predict specific phenomena that can be observed in real, complex systems. Indeed, such systems often break the basic assumptions of the TCM through the existence of multiple cavity modes, more than two levels in the matter subsystem, or by exhibiting interactions between TLSs. For example, in dielectric Whispering Gallery Mode (WGM) resonators with high dopant concentration, all of these assumptions could be broken. In such a system, WGMs of a macroscopic cylindrical crystal resonator are coupled to impurity ions within the resonator crystal lattice.\cite{PhysRevB.88.224426,goryachev2013giant,goryachev2014controlling}

In general, the regime of wave-matter interaction is primarily described by three parameters: the coupling strength $g$, the electron spin resonance (ESR) linewidth $\lambda_\text{ESR}$, and the cavity linewidth $\lambda_\text{cav}$. The strong coupling regime of this interaction is achieved when 
\begin{equation}
\color{black}
\hbox{$g>\tfrac{1}{2}\left(\lambda_\text{ESR}+\lambda_\text{cav}\right)$}.
\label{eq:strcoupcondition}
\color{black}
\end{equation}
 Because $g\propto\sqrt{N}$, where $N$ is the number of spins, the spin-photon coupling strength may be increased by doping the crystal, however this also increases the cavity linewidth due to increased loss.
This trade-off results in an optimal doping strength, which depends on the impurity ions in the crystal. The limiting of vanishingly low concentrations of ions, in particular naturally occurring impurities in sapphire, has been studied before\cite{PhysRevB.88.224426}. None of the observed ions, such as Cr$^{3+}$, V$^{2+}$, or Fe$^{3+}$ have achieved the strong coupling regime, regardless of the extraordinarily high quality factor of the WGMs, $Q>10^9$. This is due to the inability to control the ESR linewidth, which is marginally larger than the photon-spin coupling in these crystals. Thus, it is interesting to investigate the \ case in the opposite limit, where the crystal is highly doped, however as discussed above such systems should demonstrate features in addition to a conventional TCM system.

With respect to the previous work\cite{PhysRevB.87.094412,PhysRevB.88.224426}, an ideal candidate for an experiment in the opposing limit is a ruby crystal (Cr$^{3+}$:Al$_2$O$_3$) WGM resonator. Such devices have already been studied at liquid helium temperatures (4\:K) and microwave frequencies\cite{Hartnett_ruby} for application to frequency standards. This study revealed relatively high quality factors $Q>10^6$ and significant effects due to impurity ions. \color{black}{The spin photon interaction has been observed previously in Cr$^{3+}$:Al$_2$O$_3$ with a collective coupling $g=38$\:MHz\cite{Schuster:2010rm}}\color{black}.  The first millikelvin study of these ions in ruby WGM resonators at K$_\text{u}$ band (13.9 GHz) frequencies is presented.

\section{System Description}
 \subsection{Spin-Wave interaction in highly doped crystal resonators}

Interactions between WGMs and the Cr$^{3+}$ spin bath can be described by the Hamiltonian:
\begin{multline}
	\mathcal{H}=\sum_{j}\left(g_L\beta \textbf{B}\cdot{\textbf{S}_j}+D[{(\textbf{S}}_j^z)^2-\tfrac{1}{3}S_j(S_j+1)]\right)\\
+\sum_{i}\hbar \omega_{i}a_{i}^{\dagger}a_{i}+\sum_{ij} \tilde{g}_{ij}(S^{+}_{j}a_{i}+a_{i}^{\dagger}S_{j}^{-})\\
 +\frac{1}{2}\sum_{ij}{J}_{ij} \textbf{S}_{i}\cdot\textbf{S}_{j}
	\label{eq:fullham}
\end{multline}
where $g_L$ is the Land\'e $g$-factor, $\beta$ is the Bohr magneton, $\textbf{B}$ is the vector magnetic field strength, $a_{i}^{\dagger}$ and $a_{i}$ are the the creation and annihilation operators of distinct WGMs, $\omega_{i}$ are the corresponding angular frequencies at $B=\infty$ i.e. where all interactions with the spin bath are negligible, $\textbf{S}$ is the is the electronic spin operator where $\textbf{S}_i$ are the components of the components of the electronic spin operator, ${S}^{\pm}=S^{x}\pm \textrm{i} S^{y}$, $D$ is the (second order) zero field splitting (ZFS) parameter, and $\tilde{g}_{ij}$ is the effective coupling between the spin and WGM.

All terms in the first summation describe an uncoupled spin in an external DC magnetic field. The second summation is the collection of all WGMs at infinite magnetic field where all ions are detuned from the resonance frequencies. The third summation is the WGM-spin coupling terms, and the fourth summation describes the spin-spin interaction.

 \subsection{Physical Realization}
In previous studies of sapphire WGM resonators using this technique, the crystals used contained a concentration of approximately 0.1 parts per million (ppm) Fe$^{3+}$. The ruby crystal under study in the present work has had its concentration of Cr$^{3+}$ previously mesaured at 34 ppm\cite{Hartnett_ruby}.
The spin linewidth of Cr$^{3+}$ is 9\:MHz\cite{Hartnett_ruby}, whereas Fe$^{3+}$ is 27\:MHz\cite{PhysRevB.87.094412}.

To characterize the behaviour of the Cr$^{3+}$ ESR in ruby, the crystal is measured in transmission with a DC magnetic field applied parallel to the crystal axis. The cylindrical ruby sample under study has a diameter of $29.97$~mm and a height of $23.86$~mm, and is cut so that the $c$-axis of the ruby is parallel to the $z$-axis of the cylinder. To obtain the highest $Q$-factor of WGMs at low temperature, the ruby crystal is mounted within an OHFC copper cavity. Two straight antennae are orientated parallel to the cylindrical $z$-axis on either side of the copper cavity in order to couple to the WGH (quasi-Tranverse Electric) modes of the crystal. WGH modes couple to the ion impurities more effectively than WGE (quasi-Transverse Magnetic) since their electric field is orthogonal to the applied DC magnetic field. Since the concentration of dopant was high, computer software was not used to identify the mode patterns of WGMs in the crystal because frequencies were shifted in the region measured. The crystal was cooled using a dilution refrigerator (DR), with a cooling power of 1.5 W at the 4 K stage and 500 $\mu$W at 100 mK on the mixing chamber.

The copper cavity was suspended from the mixing chamber of the DR into the bore of a 7~T superconducting magnet via an Oxygen-free high thermal conductivity (OHFC) copper rod. Attenuators were inserted into the microwave line down into the fridge, with 10 dB attenuation on the 4 K stage, 10 dB attenuation on the 1 K stage, and 20 dB on the 20 mK stage. These attenuators were cooled to allow thermalization of the attenuation and allow low noise signals to be transmitted to the crystal. Measurements of the WGMs were performed using a network analyzer with incident power of -60 dBm on the ruby. A microwave isolator was attached to the output of the microwave cavity, and a cryogenic amplifier was placed on the output line at the 4 K stage for low noise amplification. A room temperature amplifier was used to boost the signal before returning to the Network Analyzer. The setup of this experiment is similar to that of previous works\cite{PhysRevB.88.224426,goryachev2013giant}.

\begin{figure*}
	\centering
		\includegraphics[width=.8\linewidth]{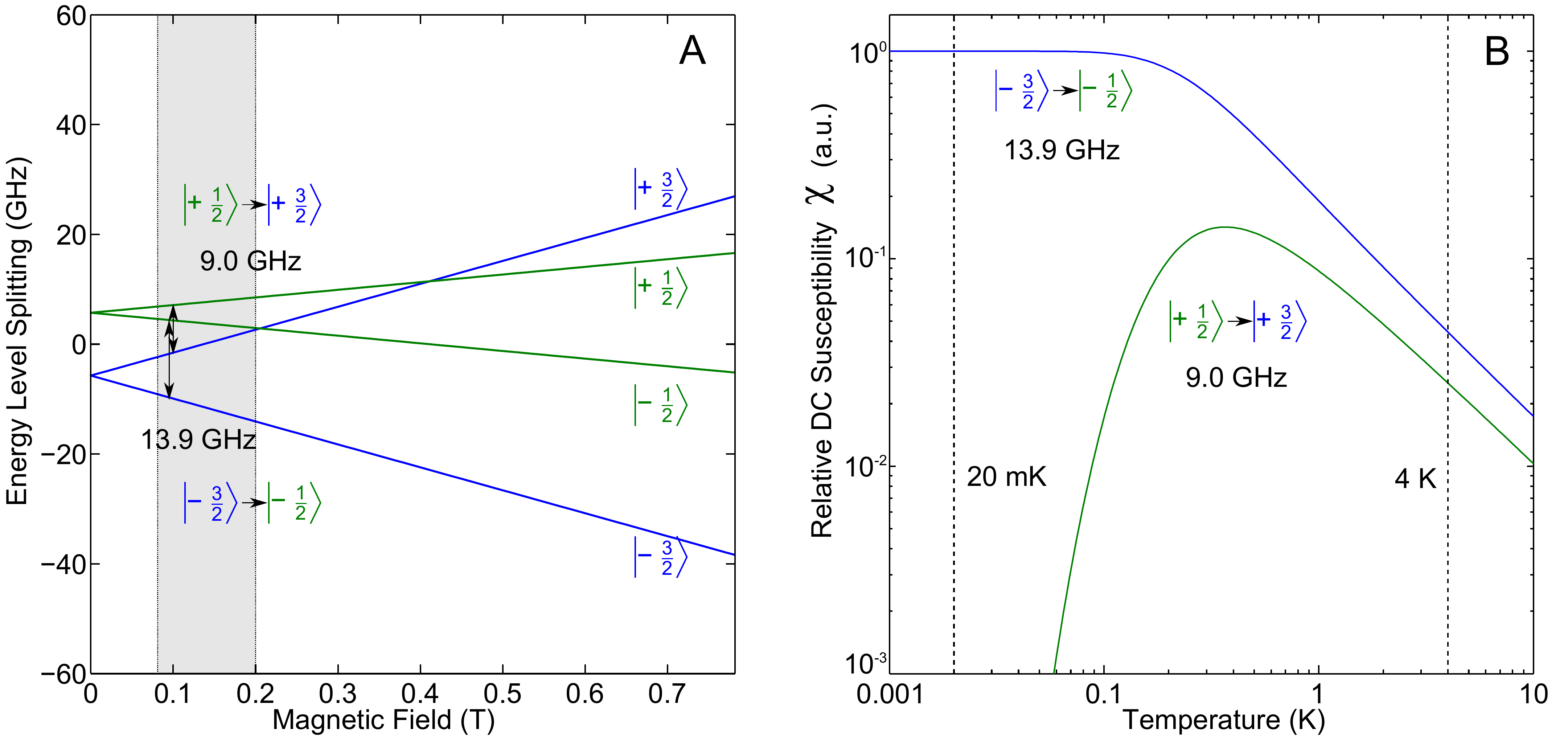}
	\caption{\color{black}(a) Energy level splitting of the Cr$^{3+}$ ion. The ground state of this system is the $\ket{-\frac{3}{2}}$ state.  (b) Relative suceptibility of the transitions as a function of temperature, with an applied DC magnetic field of 0.08 T.}
	\label{fig:hamiltoniancr3ppluschi}
\end{figure*}

The Hamiltonian of the $3d^3$ Cr$^{3+}$ ion impurity is:
\begin{align}
\mathcal{H}=&\sum_{j}\left(g_L\beta \textbf{B}\cdot{\textbf{S}_j}+D[{(\textbf{S}}_j^z)^2-\tfrac{1}{3}S_j(S_j+1)]\right),
\end{align}
with the variables as described for Eq.~\ref{eq:fullham}. For dilute Cr$^{3+}$ in sapphire\cite{PhysRev1963_132_1029_1036}, $D=-5723.5\pm 3$ MHz, and $g_{L||}$=1.984. The eigensolutions of the Hamiltonian are shown in Fig.\:\ref{fig:hamiltoniancr3ppluschi}. \color{black}The effect of a negative $D$ (ZFS parameter) is shown in Figure\:\ref{fig:hamiltoniancr3ppluschi}(a), where the $\ket{-\frac{3}{2}}$ state is the ground state. Figure\:\ref{fig:hamiltoniancr3ppluschi}(b) shows the relative susceptibilites of the two transitions. The suceptibility of ESR interactions has a strong dependence on temperature\cite{PhysRevB.89.224407}. At temperatures below $\sim$300 mK, electrons in the $\ket{+\frac{3}{2}}$ energy state will start condense to the $\ket{+\frac{3}{2}}$ ground state. As a result, the $\ket{-\frac{3}{2}}\rightarrow\ket{-\frac{1}{2}}$ transition of Cr$^{3+}$ becomes stronger and has more influence on WGMs. The microwave field is swept over a range of 72 MHz at around 13.9 GHz as the DC magnetic field is increased.

\color{black}

\section{Spin-WGM interaction}

\begin{figure*}[htpb]
	\centering
		\includegraphics[width=.72\linewidth]{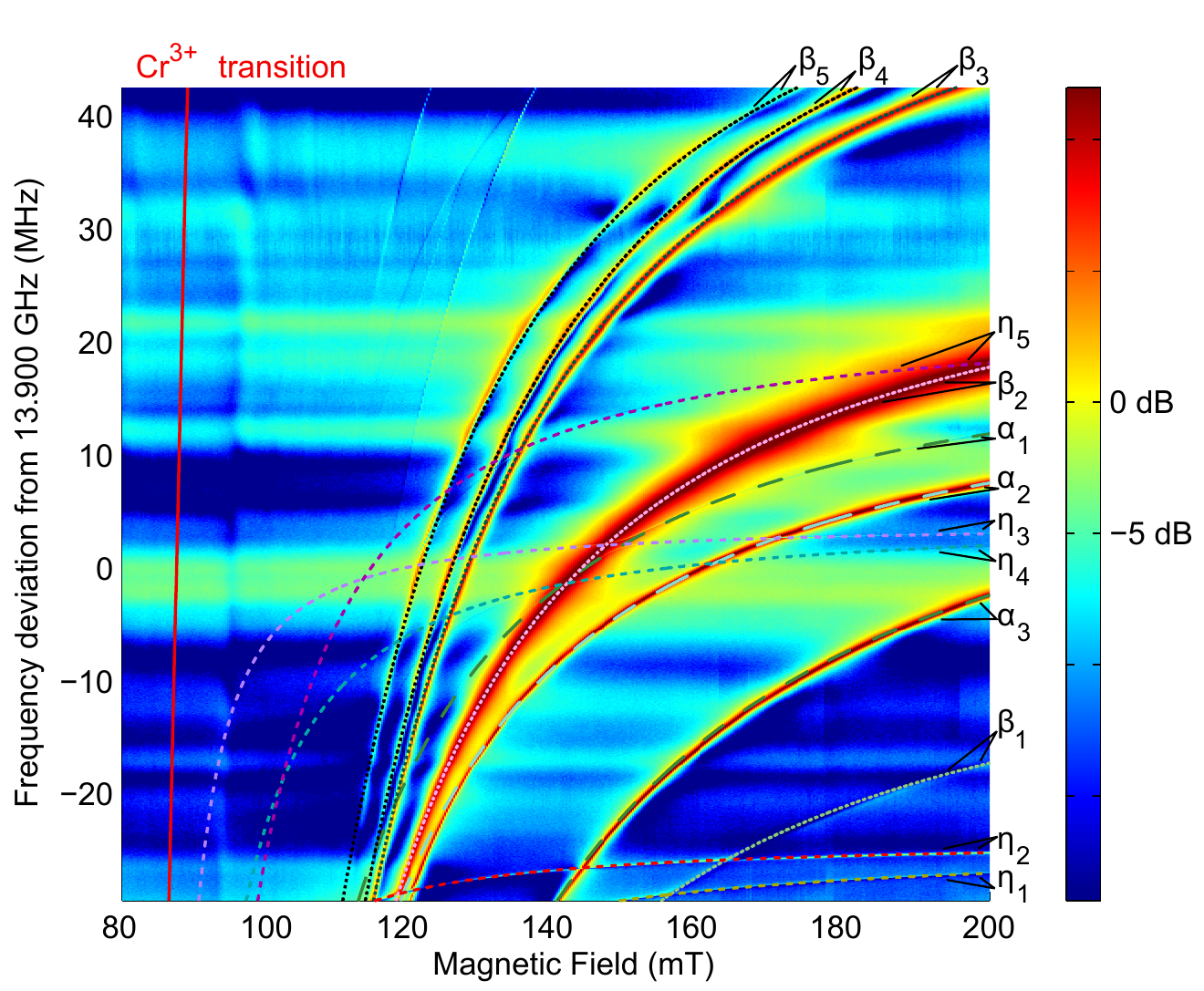}
	\caption{Composite image of a series of measurements of the power transmission (S$_{21}$) of the reuby resonator, at frequencies around 13.9 GHz at a minimum temperature approximately 20 mK, The plot shows a number of WGMs experiencing different susceptibilities due to the Cr$^{3+}$ ESR. The line representing the bare ESR transition is also shown}
	\label{fig:modesat1395RUBY_}
	
\end{figure*}

Since the system is probed in a narrow frequency band of 72 MHz, the spin system may be considered as a two level system, such that the zero field splitting (ZFS) term from Eq.~\ref{eq:fullham} may be neglected.
Next, only the WGM eigensolutions are considered within our 72 MHz span, which is a total of 18 modes (see Table \ref{table:plotcouplings1}). Because of the low power of the system, only the WGM-spin interaction is considered, and the spin-spin interaction is ignored such that the Hamiltonian becomes:
\begin{multline}
\mathcal{H}=\sum_{i}\hbar \omega_{{i}}a_{i}^{\dagger}a_{i}+\frac{\hbar}{2}\omega\sum_j\sigma_j^z
+\sum_{ij}k_{ij}(\sigma^{+}_{j}a_{i}+a_{i}^{\dagger}\sigma_{j}^{-})
\end{multline}
where $\omega(B) =\frac{2g_L\beta B}{\hbar}$ is the ESR frequency, and
\begin{align}
\sigma_j^z=&\ket{-\tfrac{1}{2}}\bra{-\tfrac{1}{2}}-\ket{-\tfrac{3}{2}}\bra{-\tfrac{3}{2}}\\
\sigma_j^+=&\ket{-\tfrac{1}{2}}\bra{-\tfrac{3}{2}}\\
\sigma_j^-=&\ket{-\tfrac{3}{2}}\bra{-\tfrac{1}{2}}
\end{align}
are the spin operators.
By assuming the WGMs to be independent, the Hamiltonian becomes the Tavis-Cummings Hamiltonian. This may then be considered as a coupled two-resonator model, with the ESR and WGM mode frequencies modelled as:  
\begin{multline}
\displaystyle \left(\frac{f_{\pm}(B)+f_{\alpha}}{f_{\alpha}}\right)^2=\\ 2 + \rho(2 + \rho) 
\pm \sqrt{\rho^2\left(2 + \rho\right)^2 + 4g^2\left(1 +\rho\right)^2}
\label{eq:modelWS}
\end{multline}		
where $f_{\pm}$ are the frequencies of the two hybrid coupled modes,  $f_{\alpha}$ is the frequency of the bare WGM with no influence of the ESR, $g$ is the spin-photon coupling, and $\rho(B) =\frac{g_L\beta B}{f_{\alpha}\hbar}$. This model is used to fit to the WGMs in Fig. \ref{fig:modesat1395RUBY_}. \color{black}Strictly speaking, the WGMs are coupled to the ESR ensemble and are therefore hybridized modes, however for simplicity we continue to refer to them as WGMs.\color{black}

\begin{table}[ht] 
\centering 
\begin{tabular}{|c|l|l|r|c|} 
\hline 
Mode&$f_{B=0.2\text{ T}}$ (MHz)& $f_{(B=\infty)}$ (MHz) & $Q_{(B=0.2 \text{ T})}$ & 2$g$ (MHz)\\
\hline 

$\alpha_{1}$&13,911.919   &13,920.128   &260,000&411\\
$\alpha_{2}$&13,907.574   &13,917.272   &120,000&447\\
$\alpha_{3}$&13,897.620   &13,914.371   &120,000&590\\
\hline
$\beta_{1}$&13,882.761   &13,895.427   &23,000&515\\
$\beta_{2}$&13,917.833   &13,930.005   &10,000&500\\
$\beta_{3}$&13,943.564   *&13,960.214   &*23,000&580\\
$\beta_{4}$&13,947.133   *&13,963.689   &*32,000&578\\
$\beta_{5}$&13,948.937   *&13,963.322   &*13,000&539\\[1ex]
\hline 
$\eta_{1}$&13,874.837   &13,875.754   &200,000&139\\
$\eta_{2}$&13,873.012   &13,875.020   &130,000&206\\
$\eta_{3}$&13,903.070   &13,903.835   &140,000&126\\
$\eta_{4}$&13,902.019   &13,904.097   &160,000&207\\
$\eta_{5}$&13,918.197   &13,921.926   &160,000&276\\[1ex]\hline

\end{tabular} 
\label{table:plotcouplings1} 
\caption{List of WGMs studied within the measured frequency band, with full width coupling to the Cr$^{3+}$ ESR.  Not all modes existed within the 72 MHz span at 0.2T, the $f_{(B=0.2 \text{ T})}$ is estimated by fitting to the model, $Q_{(B=0.2 \text{ T})}$ is measured at the highest absolute field where the mode exists.} 
\end{table} 

The large quantity of WGMs in this region makes it difficult to identify the specific mode pattern through computer modelling that conventionally is used to identify WGMs. The modes are categorized by high and low quality factor, $Q$ values, and estimated coupling to the Cr$^{3+}$ spin resonance, listed in Table~\ref{table:plotcouplings1}. Resonances are categorized as $\alpha_i$ or $\beta_i$ if they have a coupling to the ESR over 400 MHz, modes with a $Q$ factor of at least 60,000 are labeled $\alpha_i$ otherwise they are categorized as $\beta_i$. Modes with a coupling below 300 MHz are categorized as $\eta_i$. The coupling is estimated from the model (Eq.\:\ref{eq:modelWS}).  

\begin{figure}[htpb]
	\centering
		\includegraphics[width=.95\linewidth]{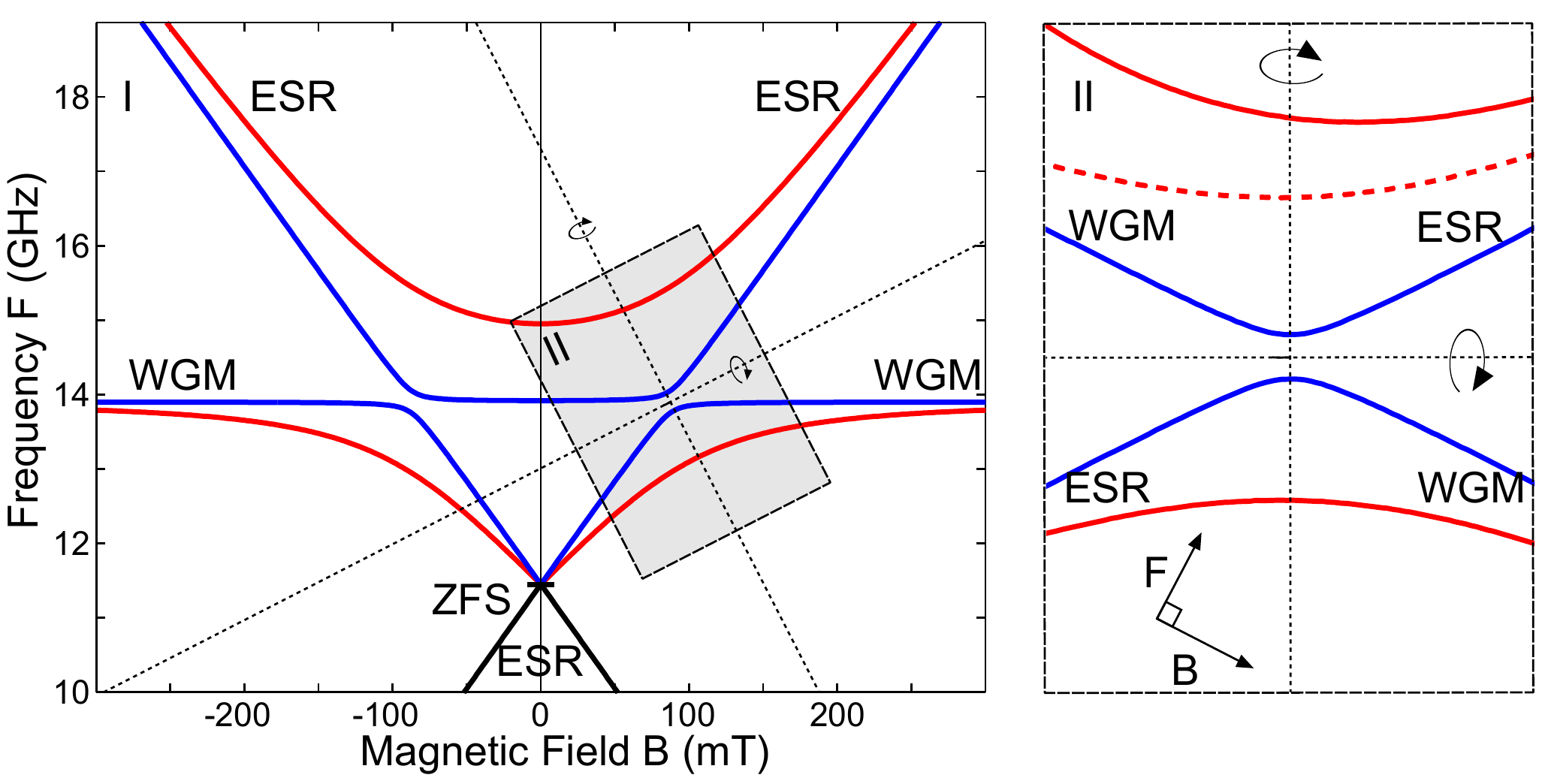}
	\caption{Model of the coupling between a WGM at 13.9 GHz (Eq. \ref{eq:modelWS}), and Cr$^{3+}$ ESR. The case of low and high coupling are shown in blue and red respectively. In the low coupling case the two modes are symmetric about the two axes around the interaction. In the high coupling case, the two modes are not symmetric due to the influence of the anti-parallel ESR transition. The shaded area in the sub-figure labelled I is rotated and presented as sub-figure II. The dashed red line shows the expected result in the case of high coupling which is symmetric about both axes, rather than the non-symmetric result observed (solid red line).}
	\label{fig:explanationcouplingexample}
\end{figure}
Figure \:\ref{fig:modesat1395RUBY_} shows the power transmitted through the ruby cavity as a function of magnetic field, with each WGM aligned to the model described in Eq.\:\ref{eq:modelWS}. The WGMs in the frequency range probed are shown in Table~\ref{table:plotcouplings1}. Some modes, for example $\eta_3$ and $\eta_4$ are not well suited to the model when fitted here at higher magnetic fields. The fit is poor at lower magnetic fields, which makes the estimated value of coupling to the ESR an underestimate, which may be due to terms in the Hamiltonian that have been ignored.
\color{black}Figure \:\ref{fig:explanationcouplingexample} demonstrates the asymmetry in the two-mode model used to estimate the spin-photon coupling. The blue line represents a low coupling case. As presented, rotated in the inset figure, it has symmetry around a pair of orthogonal axes. The highly coupled case does not have the same symmetries. This suggests that in our two mode model, the WGM is coupled to both the $\ket{-\frac{3}{2}}\rightarrow\ket{-\frac{1}{2}}$ transition, and to a lesser extent the $\ket{+\frac{3}{2}}\rightarrow\ket{+\frac{1}{2}}$ transition.

\color{black}

\section{Interaction between Whispering Gallery modes through the spin bath}

\begin{figure}[t]
	\centering
		\includegraphics[width=.95\linewidth]{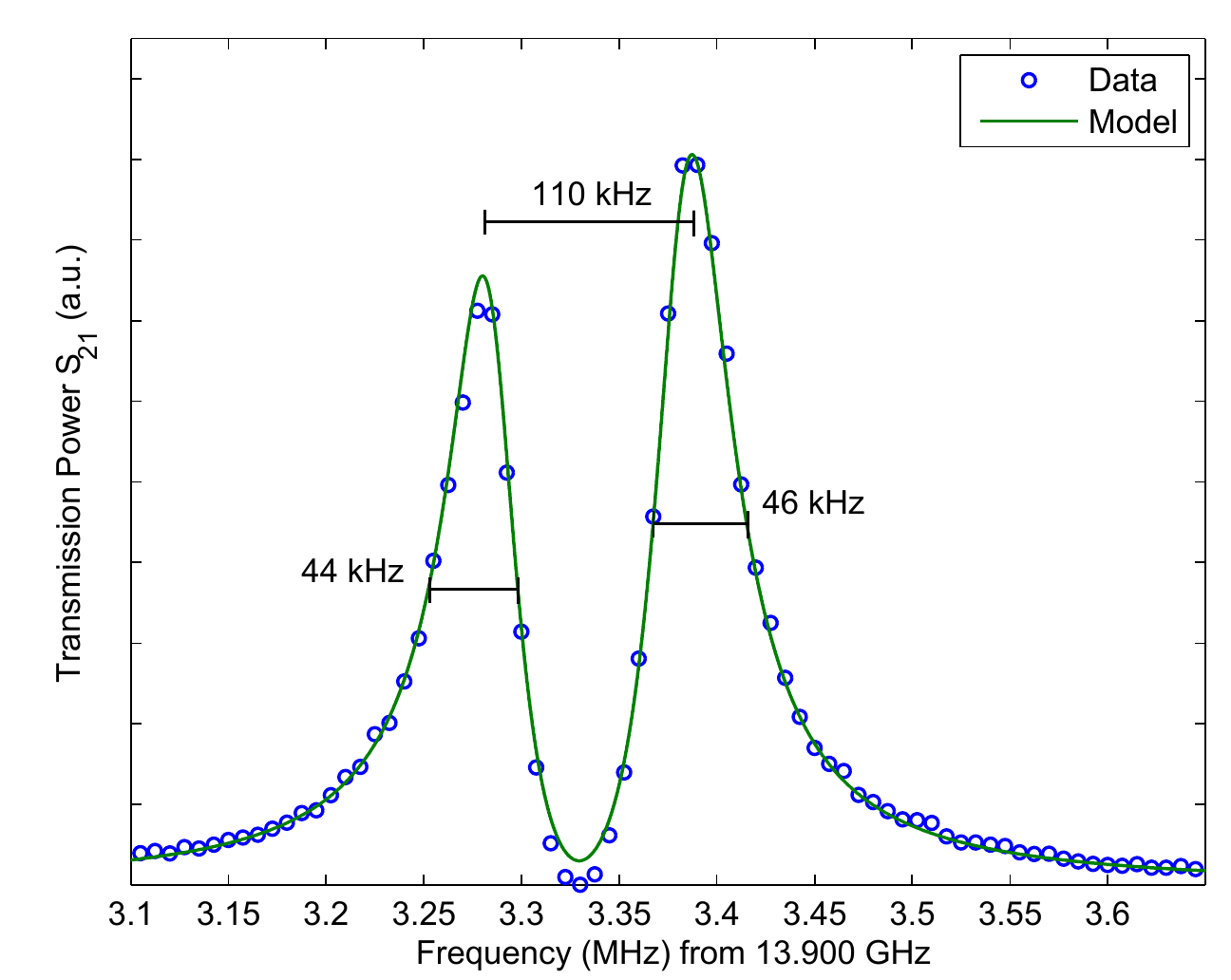}
	\caption{Power transmission of the strong coupling between the hybridized $\alpha_3$ and $\eta_3$ WGMs in Fig.~\ref{fig:avoidedcrossingsRUBY_set_of_4a}(B). The trace is calculated from the model Eq. \ref{eq:2wgmodel}.}
	\label{fig:stronginteractionruby_trace}
\end{figure}

\begin{figure*}
	\centering
		\includegraphics[width=.95\linewidth]{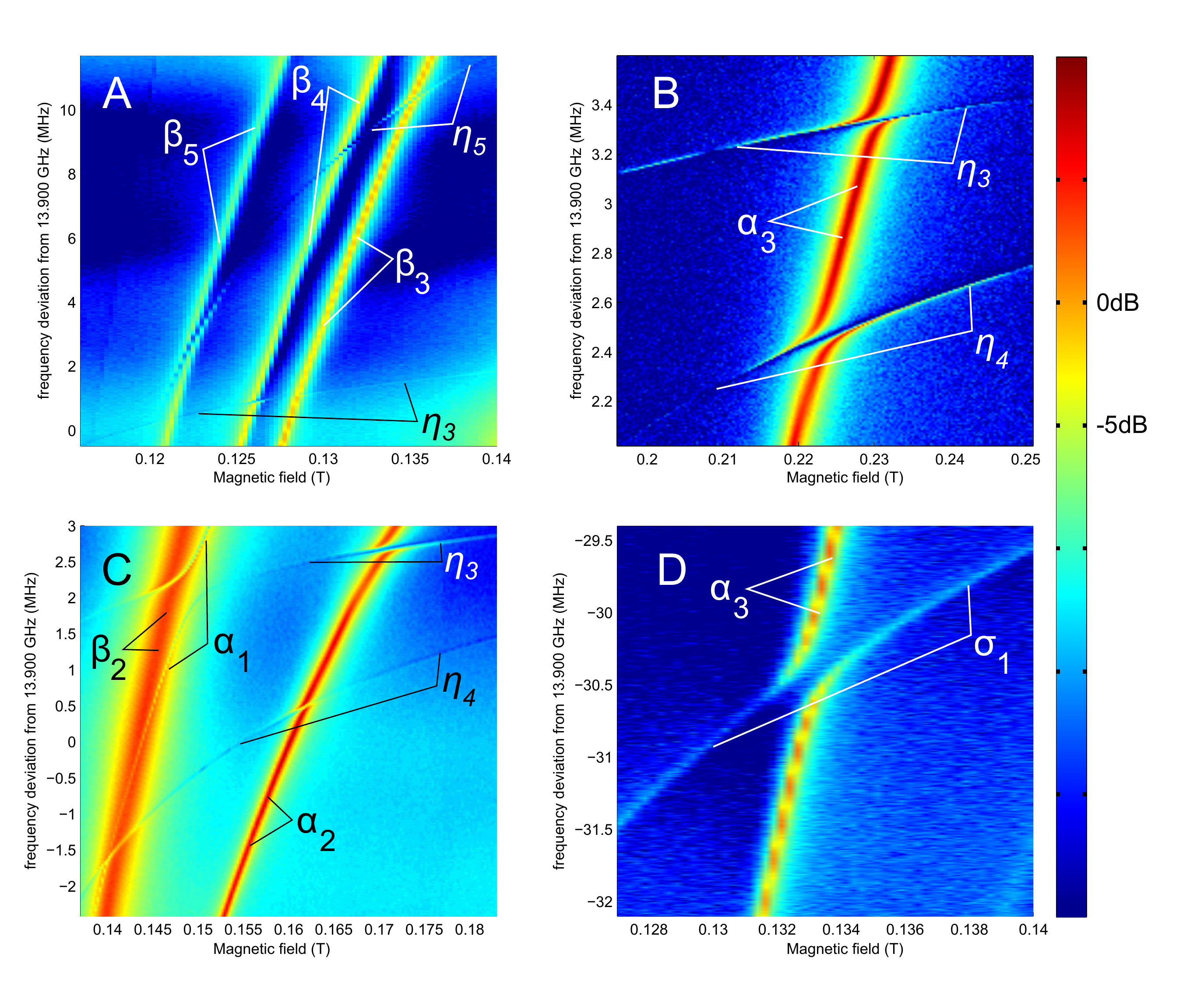}
	\caption{Series of measurements around 13.9 GHz showing avoided crossings between different WGMs with different susceptibility to Cr$^{3+}$. The difference in susceptibility causes the modes to tune across each other. Plots A and C are a subset of the measurements of Fig.~\ref{fig:modesat1395RUBY_} at 20mK, plots B and D were measured at a slightly higher temperature of 40 mK, which may result a marginally weaker coupling to the ESR.  }
	\label{fig:avoidedcrossingsRUBY_set_of_4a}
\end{figure*}

In the previous section the effect of the interaction between the electron spin and the WGM was discussed. The modes were categorized as $\alpha$ and $\beta$ for modes with high coupling to the ESR, and $\eta$ for modes with low coupling. It is in the nature of these hybrid modes of low and high suceptibility, that they will be tuned across each other as the magnetic field is tuned. In Fig.~\ref{fig:modesat1395RUBY_}, numerous modes can be observed that cross, with a selection of these crossings presented in greater detail in Fig.~\ref{fig:avoidedcrossingsRUBY_set_of_4a}. To model these interactions, the Hamiltonian (Eq.~\ref{eq:fullham}) may now be \color{black}reconsidered   as an interaction between two distinct WGM resonances. \color{black} The avoided crossing is a result of the other terms in the Hamiltonian, which breaks down to:
\begin{multline}
\mathcal{H}=\hbar \omega_{1}a_{1}^{\dagger}a_{1}+\hbar \omega_{2}a_{2}^{\dagger}a_{2}+\frac{\hbar}{2}\omega\sum_j\sigma_j^z \\
+\hbar\sum_{ij}k_{ij}(\sigma^{+}_{j}a_{i}+a_{i}^{\dagger}\sigma_{j}^{-})  +\frac{\hbar}{2}\sum_{ij}{J}_{ij} \sigma_{i}\sigma_{j}
\end{multline}
This Hamiltonian describes the coupling between 3 coupled resonators, the spin bath\cite{spinbathreview}, and a pair of WGMs.  One can simplify this by treating the system as two modes being tuned across another.
By applying a diagonalization to the Hamiltonian, we may consider the result of two harmonic oscillators, which gives a formula that approximates the transmitted power as a function of the parameters of individual modes:
\begin{multline}
P(f)=\\
\frac{k^2 (\kappa_{a}^2 + (\Delta - \delta)^2)}{\Gamma^4 + 2 \Gamma^2 (\delta\Delta -\Delta^2 + 
\kappa_{a} \kappa_{b}) + (\kappa_{a}^2 + (\Delta - \delta)^2) (\Delta^2 + \kappa_{b}^2)}
\label{eq:2wgmodel}
\end{multline}
where $k$ is the transmitted power parameter, $\Gamma$ is the coupling between the WGMs, $\kappa_{a},\kappa_{b}$ are the linewidths of the two WGMs, $\gamma=\frac{1}{2}(\kappa_{a}+\kappa_{b})$ is the mean WGM linewdth, $\Delta=f-f_0$, where $f_0$ is the frequency of the avoided crossing, and $\delta$ is the detuning of the system from the avoided crossing. In Fig.~\ref{fig:stronginteractionruby_trace}, the interaction parameters of the $\alpha_3$ and $\eta_3$ modes are measured with this model.
These  WGM interactions are listed in Table~\ref{tab:tableOfInteractions}. 
The $\alpha{_1},\eta{_3}$ interaction is the strongest, with a ratio of mean WGM linewidth to cross coupling $\frac{\gamma}{\Gamma}=8.7$. This can be categorized as strong coupling between WGMs. A similar form of strong coupling between WGMs has been observed before\cite{goryachev2014controlling} between a WGM doublet.  In the present case the interacting modes are not likely to be WGM doublets because they do not have the same frequency when $B\gg 0$.

\begin{table}[t]
	\centering
		\begin{tabular}{|l|l|r|r|l|l|l|}
		\hline
		Plot&Modes&B (mT)&$f$ (GHz)&$2\gamma$ (kHz)&$2G$ (kHz)&$\frac{G}{\gamma}$\\\hline
		A&$\beta{_4},\eta{_5}$&131.00&13.908662						&\hspace{1ex}260&\hspace{1ex}320&\hspace{1ex}1.2\\
		B$\dagger$&$\alpha_{3},\eta{_3}$&230.10&13.903339&\hspace{2ex}90&\hspace{1ex}110&\hspace{1ex}1.2\\
		B$\dagger$&$\alpha_{3},\eta_{4}$&223.20&13.902471&\hspace{1ex}100&\hspace{1ex}180&\hspace{1ex}1.8\\
		C&$\alpha{_1},\eta{_3}$&148.75&13.902063					&\hspace{2ex}75&\hspace{1ex}650&\hspace{1ex}8.7\\
		C&$\alpha{_2},\eta{_3}$&170.75&13.902693					&\hspace{1ex}108&\hspace{1ex}220&\hspace{1ex}2.0\\
		D$\dagger$&$\alpha_{3},\eta{_1}$&133.00&13.869487&\hspace{1ex}126&\hspace{1ex}550&\hspace{1ex}4.4\\\hline
		\end{tabular}
	\caption{Table of interactions. $\dagger$ denotes a measurement not made at minimum temperature, $\gamma$ is the mean linewidth of the two WGMs, and $G$ is the coupling between them}
	\label{tab:tableOfInteractions}
\end{table}

\section{Conclusion}
The coupling of WGMs to Cr$^{3+}$ spins in a ruby sample at 20 mK has been measured, with a coupling strength on the order of $g_c=$ 610 MHz. This satisfies the previous condition Eq.(\ref{eq:strcoupcondition})
 for strong coupling, as $g_c > \lambda_\text{cav}\approx$ 0.1 MHz, $\lambda_\text{ESR}=9$ MHz. This coupling strength exceeds by more than an order of magnitude the 38 MHz achieved by Schuster et al.\cite{Schuster:2010rm} on a Cr$^{3+}$:Al$_2$O$_3$ chip. However, in the present work the WGM couples to more than one ESR transition, violating an assumption of the TCM that one transition exists per system.  Thus, it would be incorrect to label this observation `true' strong coupling between cavity modes and the ion ensemble.  For WGMs that tune at different rates with magnetic field, it was found that WGM crossings occur. Some such crossings exhibited strong coupling, having a photon-photon coupling to linewidth ratio of up to 8.7.\\

\section*{Acknowledgements}
This work was funded by the Australian Research Council under grant numbers FL0992016 and CE11E0082.

\bibliography{RUBY_PAPER_final_revised}

\end{document}